\documentclass[twocolumn,showpacs,preprintnumbers,amsmath,amssymb]{revtex4}

\usepackage{graphicx,amsmath}
\usepackage{dcolumn}
\usepackage{bm}

\begin{document}

\preprint{APS/123-QED}

\title{Stress partition and micro-structure in size-segregating granular flows}%

\author{L. Staron$^{1,2,3}$} 
\email{lydie.staron@upmc.fr}
\author{J. C. Phillips$^{3}$}

\affiliation{%
$^1$ Sorbonne Univ, UPMC Univ Paris 06, UMR 7190, Inst Jean Le Rond dAlembert, F-75005, Paris, France\\
$^2$ CNRS, UMR 7190, Inst Jean Le Rond dAlembert, F-75005, Paris, France\\
$^3$ School of Earth Sciences,  University of Bristol,  United Kingdom.\\
}%

\date{\today}

\begin{abstract}
When a granular mixture involving grains of different sizes is shaken, sheared, mixed, or left to flow, grains tend to separate by sizes in a process known as size segregation. 
 In this study, we explore the size segregation mechanism in granular chute flows in terms of  pressure distribution and granular micro-structure. Therefore, 2D discrete numerical simulations of bi-disperse granular chute flows are systematically analysed.
Based on the theoretical models by Gray $\&$ Thornton \cite{gray05} and Hill $\&$ Tan \cite{hill14}, we explore the stress partition in  the phases of small and large grains, discriminating between contact stresses and kinetic stresses. Our results support both gravity-induced  and shear-gradient-induced  segregation mechanisms. However, we show that the contact stress partition is extremely sensitive to the definition of the partial stress tensors, and more specifically, to the way mixed contacts ({\it i.e.} involving a small grain and a large grain) are handled, making conclusions on gravity-induced segregation uncertain.
  By contrast, the computation of the partial kinetic stress tensors is robust. Kinetic pressure partition exhibits a deviation from continuum mixture theory of a significantly larger amplitude than contact pressure, and display a clear dependence on the flow dynamics. Finally, using a simple approximation for the contact partial stress tensors, we investigate how contact stress partition relates with the flow micro-structure,  and suggest that the latter may provide an interesting proxy for studying gravity-induced segregation.
\end{abstract}

\maketitle

\section{Introduction}

When a granular mixture involving grains of different sizes is shaken, sheared, mixed, or left to flow, grains tend to separate by sizes in a process known as size segregation. In the simple case of a free surface flow under gravity, larger grains rise to the surface while the smaller grains sink at the bottom of the flow.  Segregation occurs in a wide variety of contexts - either industrial or natural \cite{bridgewater85,powell98,felix04,frey09,rowley11,johnson12} - for which it represents an   engineering and scientific challenge. Kinetic theory offers a well established framework to describe segregation in dilute granular systems \cite{jenkins83,jenkins02}; however its applicability is limited in the case of dense flows, where grains interact through long-lasting contacts rather than binary collisions. Since the seminal work by Savage and Lun  \cite{savage88} and the introduction of the {\em random kinetic sieve}, much progress has been achieved in understanding segregation mechanisms, both experimentally \cite{berton03,phillips06,schroter06,goujon07,golick09,may10,moro10,wiederseiner11,garzo12,degaetano13,guillard13,guillard14,vaart15} and using numerical simulations  \cite{taberlet04,linares07,rognon07,yohannes10,fan11,tripathi11,marks12,thornton12,staron14,hill14}. In this respect, because they allow for very well controlled configurations and give access to the inner structure of the flow, discrete numerical simulations have proven a fruitful tool, and provide significant insight to inform continuum modeling  \cite{gray05,gray06,gray10,meruane12,thornton12,johnson12}.\\
One of the challenge posed by granular size segregation lies in the  task of identifying the actual driving mechanisms \cite{gray05,fan11,fan11b,weinhart13, hill14,tunuguntla14}. 
 Experimental evidence for the lift force that drives an intruder upwards in a quasi-static shear flow was recently given and shown to scale like Archimedes force \cite{guillard14}. It seems reasonable to suppose that such lift forces are at play in  size-disperse dense granular flows, although the dependences on the flow composition and dynamics have to be established. In the absence of an explicit generic formulation for the segregation force, a fruitful hypothesis was proposed  by Gray $\&$ Thornton \cite{gray05}: the smaller grains, while percolating through the network of larger grains, are screened from the average stress and thus carry less of the lithostatic pressure than larger grains do. This hypothetical mechanism translates mathematically into a departure from classical continuum mixture theory; it allows for the recovery of important features of segregation in depth averaged equations \cite{gray05}. It has been since the subject of various improvements \cite{gray06,marks12,thornton12,tunuguntla14}. Yet the basic ingredient, namely the fact that the partial pressure in the phase of the smaller grains is less than simply proportional to their volume occupation,  has to be unambiguously established. While  Weinhart et al do observe such asymmetry in discrete numerical simulations \cite{weinhart13}, the results obtained by Fan $\&$ Hill and  Hill $\&$ Tan are inconsistent with this hypothesis \cite{fan11,hill14}.  
 More recently, it was proposed by Fan $\&$ Hill that in addition to gravity - or in place of it - kinetic stresses play a crucial role in size segregation \cite{fan11b}. They identify the gradients of velocity fluctuations as the driving mechanism \cite{fan11,hill14}. \\
 Based on the theoretical models by Gray $\&$ Thornton \cite{gray05} and Hill $\&$ Tan \cite{hill14}, we explore the stress partition in  the phases of small and large grains, discriminating between contact stress and kinetic stress, in numerical bi-disperse granular flows. The numerical method and simulation are presented in section \ref{sec:num}.  Contact stresses  and their partition between the phases of small and large grains are analysed in section \ref{ContStress}. Section \ref{sec:kin} explores the partition of kinetic stresses and its sensitivity to the flow dynamics. The relation between contact stresses and granular micro-structure is established in section \ref{sec:micro}. We finally discuss the results in section \ref{sec:conclu}


\section{The numerical flows}
\label{sec:num}
\begin{figure}
\begin{minipage}{1.\linewidth}
\begin{minipage}{0.8\linewidth}
\centerline{\includegraphics[width = 1.\linewidth]{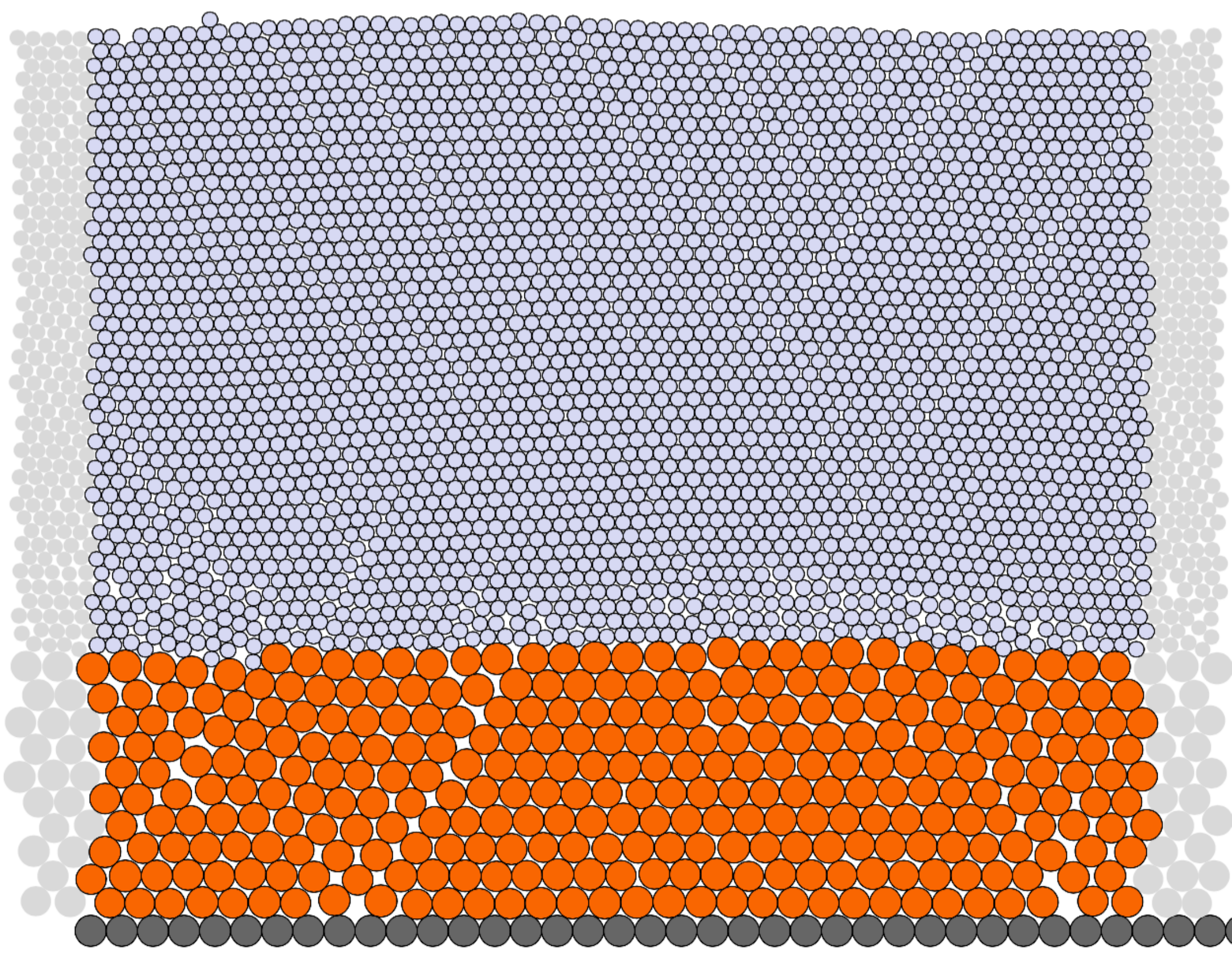}}
\end{minipage}
\vfill
\begin{minipage}{0.8\linewidth}
\centerline{\includegraphics[width = .98\linewidth]{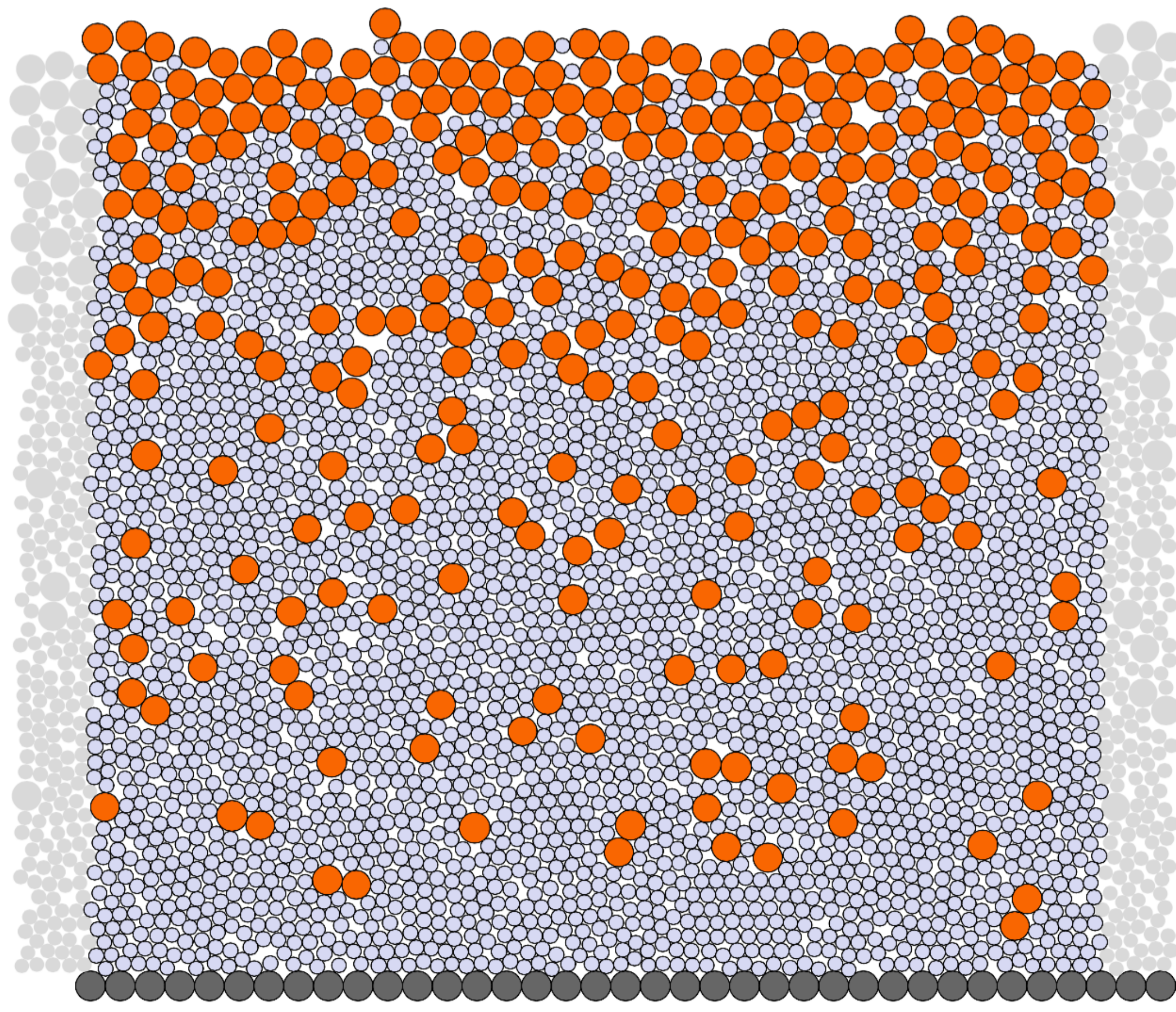}}
\end{minipage}
\caption{Example bi-disperse granular flow  with volume fraction of large beads $\Phi_L=0.3$ in the initial state ($t=0$) and after steady-state is reached and segregation occurred ($t=150s$);  the slope is $\theta=23^\circ$. The ratio between the diameter of large and small grains is $d_L/d_S = 2$ ($d_S=0.005$ m). (color online)}
\label{snaps}
\end{minipage}
\end{figure}
The numerical systems consist of two-dimensional granular beds of grains of two sizes allowed to flow under gravity on a fixed plane of slope $\theta$ made of grains of the larger size.  As the flow develops, segregation occurs, and the larger grains - placed initially at the bottom of the granular layer - rise in the flow, as illustrated in Figure~\ref{snaps}. This typical chute flow experiment is in its principle similar to those reported in \cite{rognon07,thornton12,marks12,staron14}.\\
This numerical experiment was performed using the contact dynamics algorithm  \cite{jean92,moreau94}, assuming perfectly rigid grains. The grains interact at contacts through solid friction: locally, the normal and tangential contact forces satisfy $f_t \leq \mu f_n$, where $\mu$ is the coefficient of friction at contact. Moreover, a coefficient of energy restitution $e$ sets the amount of energy dissipated by collisions. The numerical values of $\mu$ and $e$ control the effective frictional properties of the flow in a given configuration. They are strictly the same for all contacts between large and small grains. In this work, we are not interested in understanding how they may affect the segregation process, hence, their value was set to $\mu=0.5$ and $e=0.25$, and was not varied.\\
We denote  $d_L$ and $d_S$ the mean diameter of large and small grains respectively.
To prevent geometrical ordering, likely to happen for strictly mono-sized packings, both large and small grains have diameters uniformly distributed around their mean value so that  $ \frac{(d^\text{max}_{L}-d^\text{min}_{L})}{d_{L}} = \frac{(d^\text{max}_{S}-d^\text{min}_{S})}{d_{S}}=0.08$ (a discussion on the influence of this value on the segregation process can be found in \cite{staron14}).  The ratio $d_L/d_S$ was not varied: $d_L/d_S=2$.\\ 
Periodic boundary conditions were implemented to ensure long flow durations; the width of the simulation cell is $35d_L$. The basal boundary is made of a row of fixed beads of diameter $d_L$. In the initial state, a layer of large beads is overlaid by a layer of small beads (Figure~\ref{snaps}). This is achieved by random deposition under gravity. We denote $\Phi^L$ the volume fraction of large beads at the flow scale,  {\it i.e.} the ratio of the volume of large beads to the total volume of grains: $\Phi^L = V_L/(V_S + V_L)$. The volume fraction of small beads is $\Phi^S=(1-\Phi^L)$. In the simulations, $\Phi^L$ was varied between $0.06$  and $0.90$. In addition, we define the local volume fraction of large and small beads $\phi^L$ and $\phi^S$. The height of the granular bed in the initial state is $H_0$; irrespective of $\Phi^L$, $H_0$ was kept constant and equal to $H \simeq 60 d_S$. The slope of the granular bed $\theta$ was varied between $21^\circ$ and $26^\circ$, allowing different flow velocities.  The numerical values used for the simulations are the following: $d_S=5.10^{-3}$m, $\rho = 1$ kg.m$^{-2}$, $g=9.8$ m.s$^{-2}$. \\

\section{Contact stresses and gravity-driven segregation}
\label{ContStress}

\begin{figure}
\begin{minipage}{0.98\linewidth} 
\centerline{\includegraphics[angle=-90,width = 1.0\linewidth]{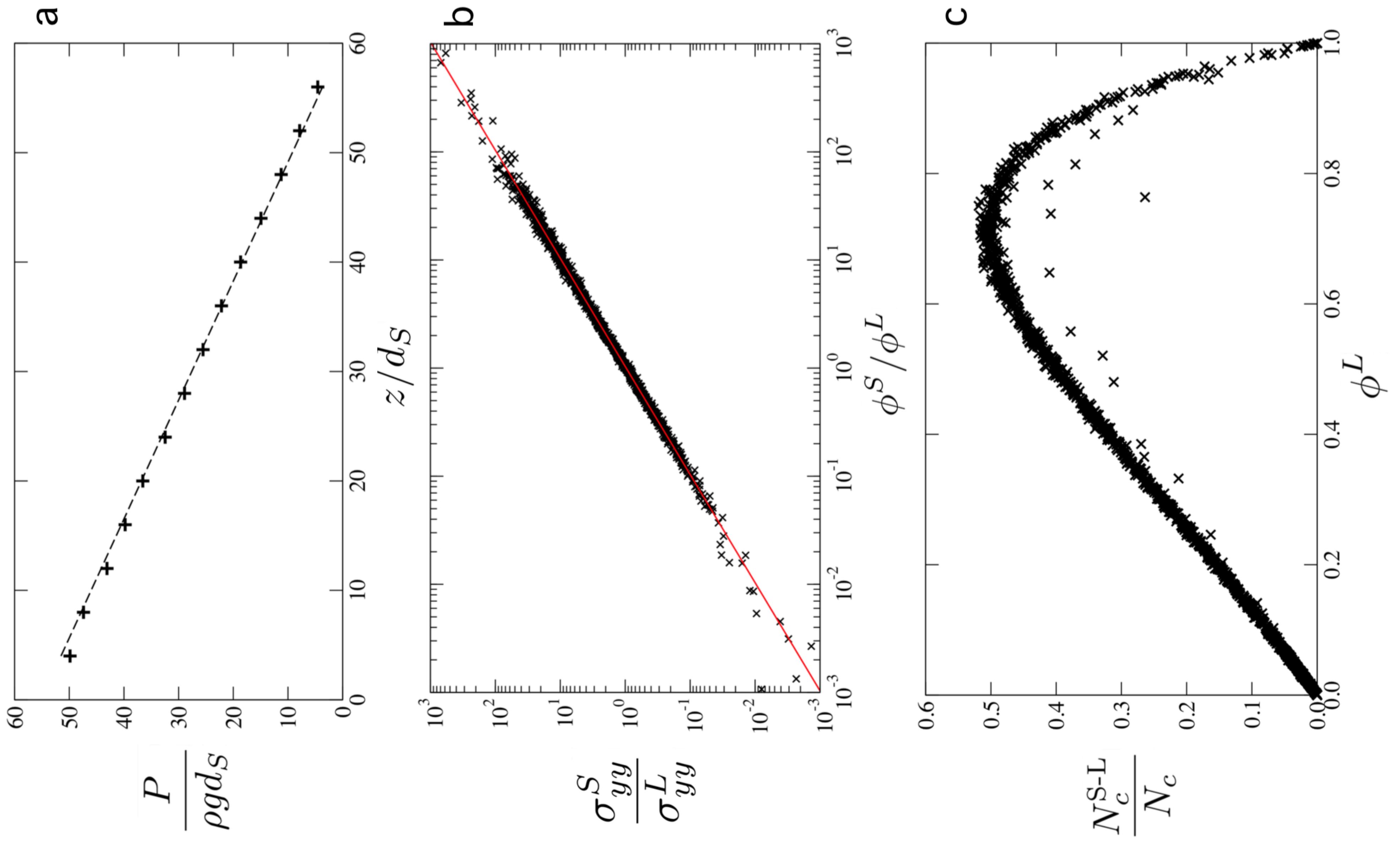}}
\caption{{\bf a} Exemple of pressure profile in a segregated flow with $\Phi^L = 0.3$ and $\theta=23^\circ$; the dotted line shows $P=\rho g (H_0-z)$; {\bf b} Ratio of the partial contact stresses  $\sigma^S_{yy}/\sigma^L_{yy}$ as a function of the local ratio of the volume fractions $\phi^S/\phi^L$ for flows with $\Phi^L=0.06$, $0.15$, $0.30$, $0.45$, $0.60$ and $0.75$,  and slope $\theta=23^\circ$; the line shows  ${\sigma^S_{yy} }/{\sigma^L_{yy} } = 0.9660 \: ({\phi^S}/{\phi^L})$;  {\bf c} Proportion of mixed contacts (between one small and one large grain) as a function of the local volume fraction of large grains for  $\Phi^L=0.06$, $0.15$, $0.30$, $0.45$, $0.60$ and $0.75$,  and slope $\theta=23^\circ$.}
\label{Partial-P}
\end{minipage}
\end{figure}

Contact stresses are transmitted at contacts between grains during either short collisions or enduring contacts; its micro-mechanical expression is the following \cite{rothenburg89}:
\begin{equation}
\boldsymbol{\sigma} = \frac1V \sum_{\alpha \in N_c} \vec{f}^\alpha \otimes \vec{\ell}^\alpha,
\label{eq:stress}
\end{equation}
where $\vec{f}^\alpha$ is the force transmitted at the contact $\alpha$, $\vec{\ell}^\alpha$ is the vector joining the centres of mass of the two grains involved, $N_c$ is the number of contacts over which the summation is made, $V$ is the volume over which the stress is computed, and $\otimes$ is the dyadic product.  The eigen values $\lambda_1$ and $\lambda_2$ of this stress tensor gives the pressure $P$: $P=(\lambda_1+\lambda_2)/2$. A typical pressure profile is shown in Figure \ref{Partial-P}a for a granular flow with $\Phi^L = 0.30$ flowing at an angle of 23$^\circ$ in the steady ({\it i.e.} segregated) state, as shown in Figure~\ref{snaps}. It simply obeys a lithostatic profile $P(z) = \rho g \cos\theta (H_0-z)$, where $z$ is the depth (counted from the bottom). \\
In a bi-disperse granular flows, contacts may involve only large grains,  only small grains, or one large grain and one small grains. In these last mixed cases,  a meaningful partition of the stress  for the computation of the partial stress tensors $\boldsymbol{\sigma}^S$ and $\boldsymbol{\sigma}^L$   is necessary. Following \cite{weinhart13,hill14}, the contribution of the mixed cases is distributed according to the size of the grains, namely weighted by a prefactor $d_L/(d_L + d_S)$ or $d_S/(d_L + d_S)$ for the phase of large and small grains respectively (we shall see that using a different partition changes the results dramatically). We thus define the partial stress tensors $\boldsymbol{\sigma}^L$ and $\boldsymbol{\sigma}^S$, and the corresponding partial pressures $P^L$ and $P^S$. In the following however, to allow precise comparison with earlier works, we will mostly use the normal stress components $\sigma^L_{yy}$ and $\sigma^S_{yy}$.\\
In the framework of classical continuum mixture theory,  partial stresses are expected to be proportional to the mean stress  $\sigma_{yy}$ and to the volume fraction of the considered grain species in the mixture, namely locally in the flow: $\sigma^L_{yy} = \phi^L\sigma_{yy}$ and $\sigma^S_{yy} = \phi^S\sigma_{yy}$. A departure from mixture theory as proposed by \cite{gray05} implies however different stress partition coefficients $\psi^L \neq \phi^L$ and $\psi^S \neq \phi^S$ \cite{fan11,weinhart13} such that  $\psi^L+ \psi^S=1$ and 
\begin{eqnarray}
\nonumber \sigma^L_{yy} &=& \psi^L\sigma_{yy},\\ 
\nonumber  \sigma^S_{yy} &=& \psi^S\sigma_{yy}. \\ \nonumber
\end{eqnarray}
Considering bi-disperse flows with a volume fraction of large grains $\Phi^L=0.06$, $0.15$, $0.30$, $0.45$, $0.60$ and $0.75$,  and flowing at an angle $\theta=23^\circ$, we evaluate the local contact stresses $\sigma^L_{yy}$ and $\sigma^S_{yy}$ using equation (\ref{eq:stress}), computed over horizontal layers of width $4d_S$, and over time intervalles of $1.25$ seconds. For the same volumes and time intervals, the local volume fraction of large and small grains $\phi^L$ and $\phi^S$ are also evaluated. We do no try to separate the early stages of the segregation from the later stages as we found that it had no visible influence on the results  shown here after.  We can thus plot the stress ratio $\sigma^S_{yy} / \sigma^L_{yy} $ as a function of volume fraction ratio $\phi^S/\phi^L$ (Figure \ref{Partial-P}b). In order to filter out the extreme cases, the best fit is calculated for $0.01<\phi^S/\phi^L  < 100$. We find 
\begin{equation}
\frac{\sigma^S_{yy} }{\sigma^L_{yy} } \simeq (0.9660 \pm 0.003) \frac{\phi^S}{\phi^L},
\label{eq:fitSig}
\end{equation}
 which suggests a  small asymmetry of the stress partition compared to a classical mixture. \\
In the original model proposed by Gray $\&$ Thornton \cite{gray05}, the stress partition obeys the following law:
\begin{eqnarray}
\label{eq:psiL} \psi^L &=& \phi^L (1 + B\phi^S) ,\\
\label{eq:psiS} \psi^S &=& \phi^S (1-B\phi^L).
\end{eqnarray}
Using (\ref{eq:fitSig}), (\ref{eq:psiL}) and  (\ref{eq:psiS})  gives $B\simeq (1-0.9660)/(\phi^L +0.9660(1-\phi^L))$, namely  $0.034 < B <0.035$. This value is larger than the value observed for 3D chute flows by Weinhart et al (who find $B\simeq 0.02$) \cite{weinhart13}, and supports the plausibility of gravity-driven segregation in the system (note that Hill $\&$ Tan find $B\simeq 0$ for 3D rotating drums \cite{hill14}).\\

It should be noted that the value of $B$ is very sensitive to the details of the stress computation. For instance, considering the ratio of the partial pressures $P^S/P^L$ rather than the ratio of the partial normal stresses $\sigma^S_{yy} /\sigma^L_{yy}$ gives $0.045 < B <0.047$, namely a larger value of $B$.  Far more dramatic is the influence of the stress distribution  at mixed contacts (between a large grain and a small grain) between the two phases. In the analysis above, the distribution of the stress transmitted at mixed contacts is weighted by the grain size, namely a prefactor  of $1/3$ for the contribution to the phase of small grains and a prefactor  of $2/3$ for the contribution to the phase of large grains (since $d_L=2d_S$). Changing the prefactors to $0.2$ for the phase of small grains and $0.8$ for the phase of large grains leads to $B\simeq 0.33$, namely a massive increase. On the contrary, changing the prefactors to $0.4$ and $0.6$ leads to $B\simeq -0.1$, that is an inverse segregation process.  Finally, changing the prefactors to $0.5$ and $0.5$ leads to $B\simeq -0.32$, and brings an entirely different picture.  The strong influence of the way mixed contacts are taken into account in the stress computation can be explained by their proportion in the mixture, which can reach locally $50\%$ depending on the volume fraction of large grains, as is visible in Figure \ref{Partial-P}c.  This makes the interpretation of contact stress ratios awkward, particularly if one wants to compare mixtures with different grain sizes. In that case, it will be difficult to tell the bias introduced by the ponderation by the grains size in the partial stresses computation from the influence on the mixed stress partition that derives.  \\
No influence of the slope angle $\theta$ (namely of the flow velocity) on the value of $B$ was observed.

\section{Kinetic stresses and shear-gradient-induced segregation}
\label{sec:kin}

\begin{figure}
\begin{minipage}{0.98\linewidth} 
\centerline{\includegraphics[angle=-90,width = 1.\linewidth]{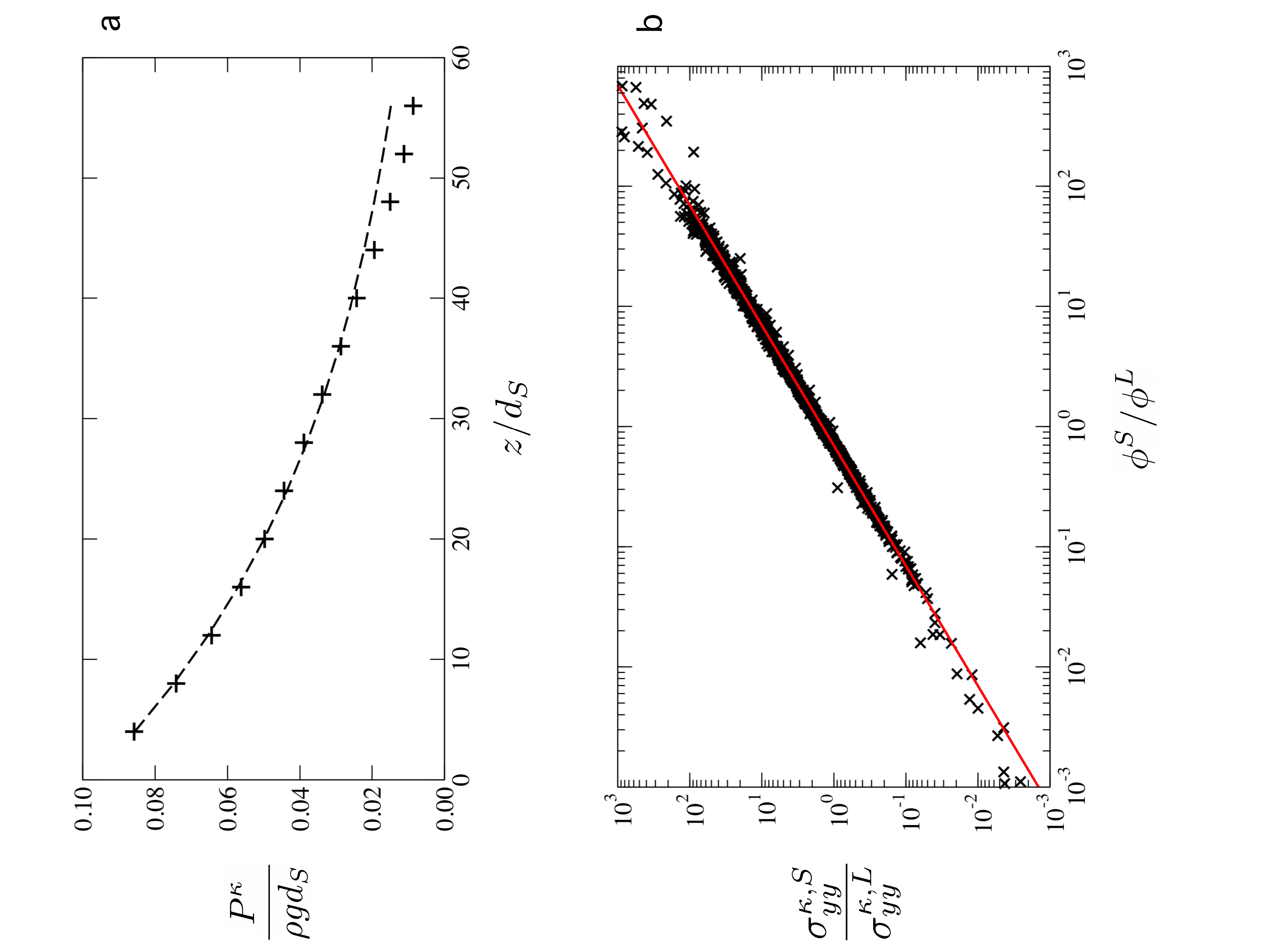}}
\caption{ {\bf a}: Profile of the kinetic pressure $P^\kappa$ (normalised by $\rho g d_S$)  for a granular flow with $\Phi^L = 0.30$ and $\theta=23^\circ$ in steady state; the dotted line shows the best fit $P^\kappa = 0.98 e^{-z/(14.8d_L)} $;  {\bf b} Ratio of the local partial kinetic stresses $\sigma^{\kappa,S}_{yy}/\sigma^{\kappa,L}_{yy}$ as a function of the ratio of the local volume fractions $\phi^S/\phi^L$ for $\Phi^L=0.06$, $0.15$, $0.30$, $0.45$, $0.60$ and $0.75$,  and slope $\theta=23^\circ$; the plain line shows the best fit $y=1.44x^{1.0057}$} 
\label{Partial-PK}
\end{minipage}
\end{figure}

\begin{figure}
\begin{minipage}{0.98\linewidth} 
\centerline{\includegraphics[angle=0,width = 1.\linewidth]{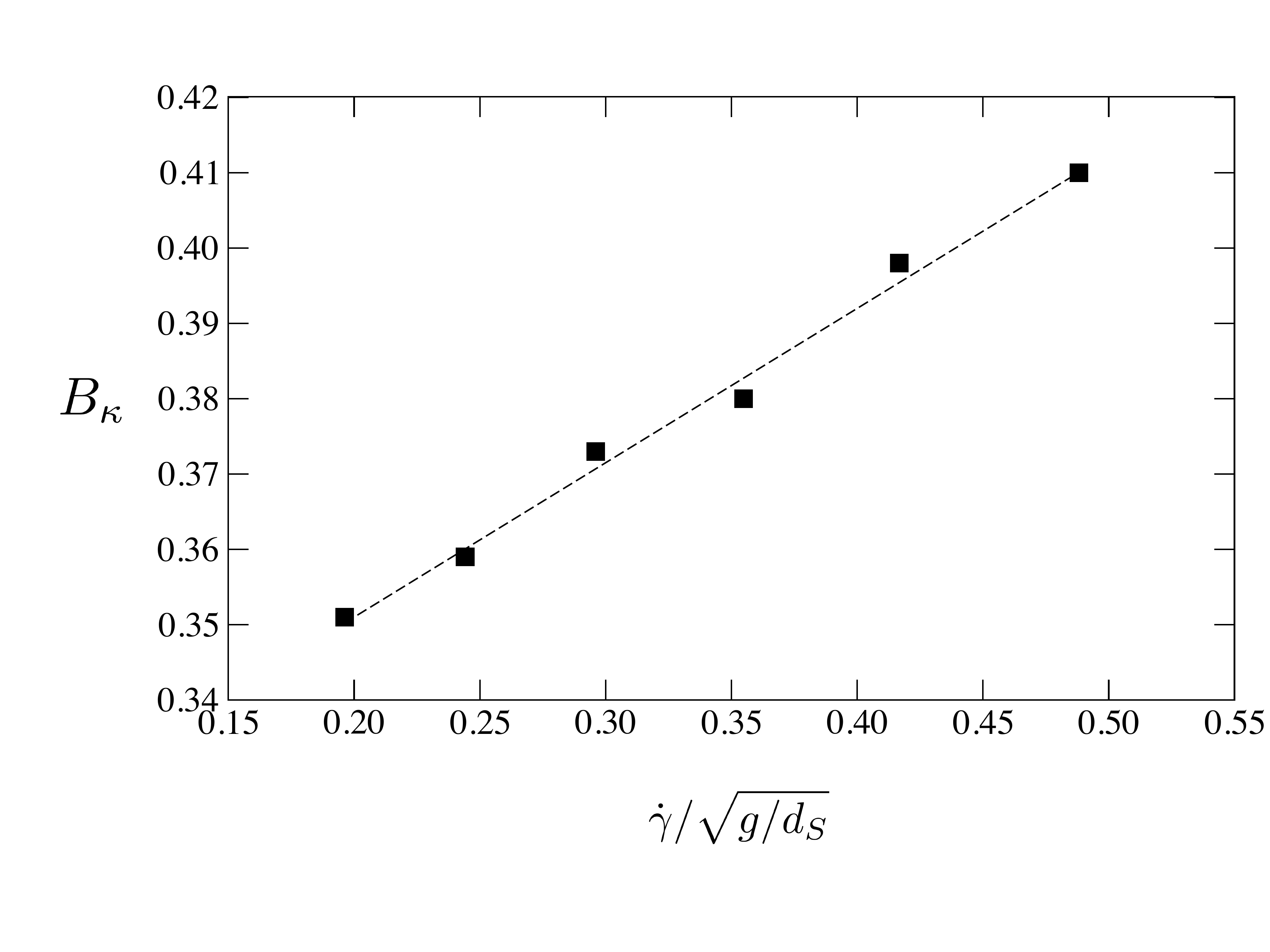}}
\caption{Kinetic pressure partition coefficient $B_\kappa$ as a function of the flow normalised mean shear rate  $\dot{\gamma}/\sqrt{g/d_S}$ computed for flows with a mean volume fraction of large grains $\Phi^L=0.45$ and for slope angles varying from $\theta=21^\circ$ to $\theta=26^\circ$. The dotted line shows the  best fit trend $y=0.310 + 0.205x$.}
\label{Bkappa}
\end{minipage}
\end{figure}

Kinetic stresses are related to the existence of velocity fluctuations in granular flows, and can be quantified through an analogue of the Reynolds stress tensor:
\begin{equation}
\boldsymbol{\sigma}^\kappa = \frac1V \sum_{i \in N_p} m_i {\delta \vec{v}}_i \otimes {\delta \vec{v}}_i,
\end{equation}
where ${\delta \vec{v}}_i$ is the fluctuating velocity of the grain $i$, $m_i$ its mass,  $N_p$ the total number of grains in the volume $V$ over which the stress is computed. A typical kinetic pressure profile is shown in Figure \ref{Partial-PK}a for a steady state granular flow with $\Phi^L = 0.30$ flowing at an angle of 23$^\circ$. The normalisation by $\rho g d_S$ allows for quantitative comparison with the contact stress profile shown in Figure \ref{Partial-P}. Kinetic stresses are about four orders of magnitude smaller than contact stresses, namely seemingly negligible. However, as stressed by \cite{fan11b,hill14}, the difference of behaviour between the phase of small grain and the phase of large grain is more striking than for contact stresses. Computing the partial kinetic stress tensors $\boldsymbol{\sigma}^{\kappa,S}$ and $\boldsymbol{\sigma}^{\kappa,L}$ is straightforward as it implies a summation on the grains and not on the contacts.
 Plotting the kinetic stress ratio $\sigma^{\kappa,S}_{yy} / \sigma^{\kappa,L}_{yy} $ as a function of volume fraction ratio $\phi^S/\phi^L$ (Figure \ref{Partial-PK}b),  we find 
$$\frac{\sigma^{\kappa,S}_{yy} }{\sigma^{\kappa,L}_{yy} } \simeq (1.4413 \pm 0.0022) \left(\frac{\phi^S}{\phi^L}\right)^{1.0057}.$$
The prefactor $1.4413$ implies that smaller grains see more of the kinetic stress than larger grains, {\it i.e.} they undergo larger velocity fluctuations. Since they are geometrically less constrained than the larger grains due to their size, this result is expected.
 Interpreting this value in the framework of the model by Gray $\&$ Thornton \cite{gray05}, namely introducing a kinetic pressure partition coefficient $B_\kappa$ such that: 
 \begin{eqnarray}
\label{bk1} \sigma^{\kappa,L}_{yy} &=& \phi^L (1 + B_\kappa \phi^S)  \sigma^\kappa_{yy},\\
\label{bk2}  \sigma^{\kappa,S}_{yy} &=& \phi^S (1 - B_\kappa \phi^L)  \sigma^\kappa_{yy},
\end{eqnarray}
  gives a large negative pressure partition coefficient  $B_\kappa \simeq -0.373$, similar to the values observed by \cite{hill14} and \cite{weinhart13} for 3D flows ($B_\kappa \simeq -0.39$ and $B_\kappa \simeq -0.38$ respectively).  The value of $B_\kappa$  is more then 10 times larger than the pressure partition coefficient measured for contact stresses in section \ref{ContStress}, and supports the proposition of a segregation mechanism driven by granular temperature by \cite{hill14}. In the chute flow configuration, large grains are segregated towards  the cooler regions of the flow (namely the surface) as predicted by kinetic theory \cite{larcher13}, but differently from what is observed in rotating drums, for which the flow surface coincides with higher granular temperature \cite{hill14}. \\
We can show that the value of $B_\kappa$ is sensitive to the mean shear rate. Simulating granular flows with $\Phi^L = 0.45$, and with varying slopes $\theta=21^\circ$, $22^\circ$, $23^\circ$, $24^\circ$, $25^\circ$, and $26^\circ$, we compute the ratio $\sigma^{\kappa,S}_{yy} / \sigma^{\kappa,L}_{yy} $ for each case,  and fit the  dependence on ${\phi^S}/{\phi^L}$ by a linear law. We then derive the mean value of $B_\kappa$ from  relations (\ref{bk1}) and  (\ref{bk2}). The plot of $B_\kappa$ as a function of the normalised shear rate is displayed in Figure \ref{Bkappa}; we observe a monotonous increase with $\dot{\gamma}/\sqrt{g/d_S}$.  From the analysis of the segregation time scales for similar 2D numerical flows \cite{staron14},  this increase of $B_\kappa$ corresponds to shorter segregation time, {\em i.e.} a greater segregation velocity. However,  it does not coincide with a larger segregation rate, as the final position of the center of gravity of the large grains was shown to  remain unaffected by increasing the mean shear rate \cite{staron14}.

\section{Micro-structural signature} 
\label{sec:micro}

While one can easily picture why smaller grains are submitted to larger kinetic stresses (simply as a result of their greater degrees of freedom), the fact that they sustain a smaller proportion of the contact stresses than their volume proportion is less intuitive.   From the micro-mechanical definition of the contact stress tensor given in (\ref{eq:stress}), we can try to approximate the partial pressures by the mean forces and the mean coordinance number in each phase of grains. \\
If $N_c^L$ is the number of contacts involving at least one large grain, $f_L$ the mean modulus of the forces transmitted by these contacts (in which a least one large grain is involved), and $\ell_L$ is the mean distance between the centres of mass of the two grains in contacts, then an estimate  $\Pi^L$ of the partial pressure $P^L$ supported by  the large grains is (following (\ref{eq:stress})):
 $$\Pi^L \simeq \frac{1}{V}N_c^L {f_L}{ \ell_L}.$$
The number of contacts $N_c^L$ involving at least one large grain  can be estimated from the number of large grains $n_L$, and their coordination number $z_L$ ({\it ie} the number of contacts that a large grain is experiencing on average): $N_L = n_L z_L/2 $.
The volume occupied by large grains is $\phi^L V$; the mean diameter of the large grains being $d_L$, we  estimate their number $n_L = \phi^L V/(\pi {d^2_L}/4)$.  Finally,  $\ell_L$ can be approximated by the diameter of one large grain $\ell_L \simeq d_L$. This gives the following estimate for the magnitude of the partial pressure supported by  large grains:
\begin{equation}
\label{eq:pL}
 \Pi^L \simeq \frac{2}{\pi d_L}{z_L f_L} \times \phi^L .
\end{equation}

In the same way, $N_c^S$ being the number of contacts involving at least one small grain, $f_S$ the mean modulus of the forces transmitted by these contacts (in which a least one small grain is involved), and $\ell_S $ the mean distance between the centres of mass of the two grains in contact, an estimate $\Pi^S$ of the partial pressure $P^S$  supported by  small  grains is:
 $$\Pi^S \simeq \frac{1}{V} N_S {f_S}{ \ell_S}.$$
 As precedently, $N_S = n_S\times z_S/2$. 
Smaller grains occupying a volume $\phi^S$,  their number is $n_S = \phi^S V/(\pi {d^2_S}/4)$. Finally, reasoning as for large grains, we suppose $\ell_S \simeq d_S $ and eventually
\begin{equation}
\label{eq:pS}
 \Pi^S \simeq   \frac{2}{\pi d_S}{z_S f_S} \times \phi^S.
 \end{equation}
\begin{figure}
\begin{minipage}{0.98\linewidth} 
\centerline{\includegraphics[angle=0,width = 1.\linewidth]{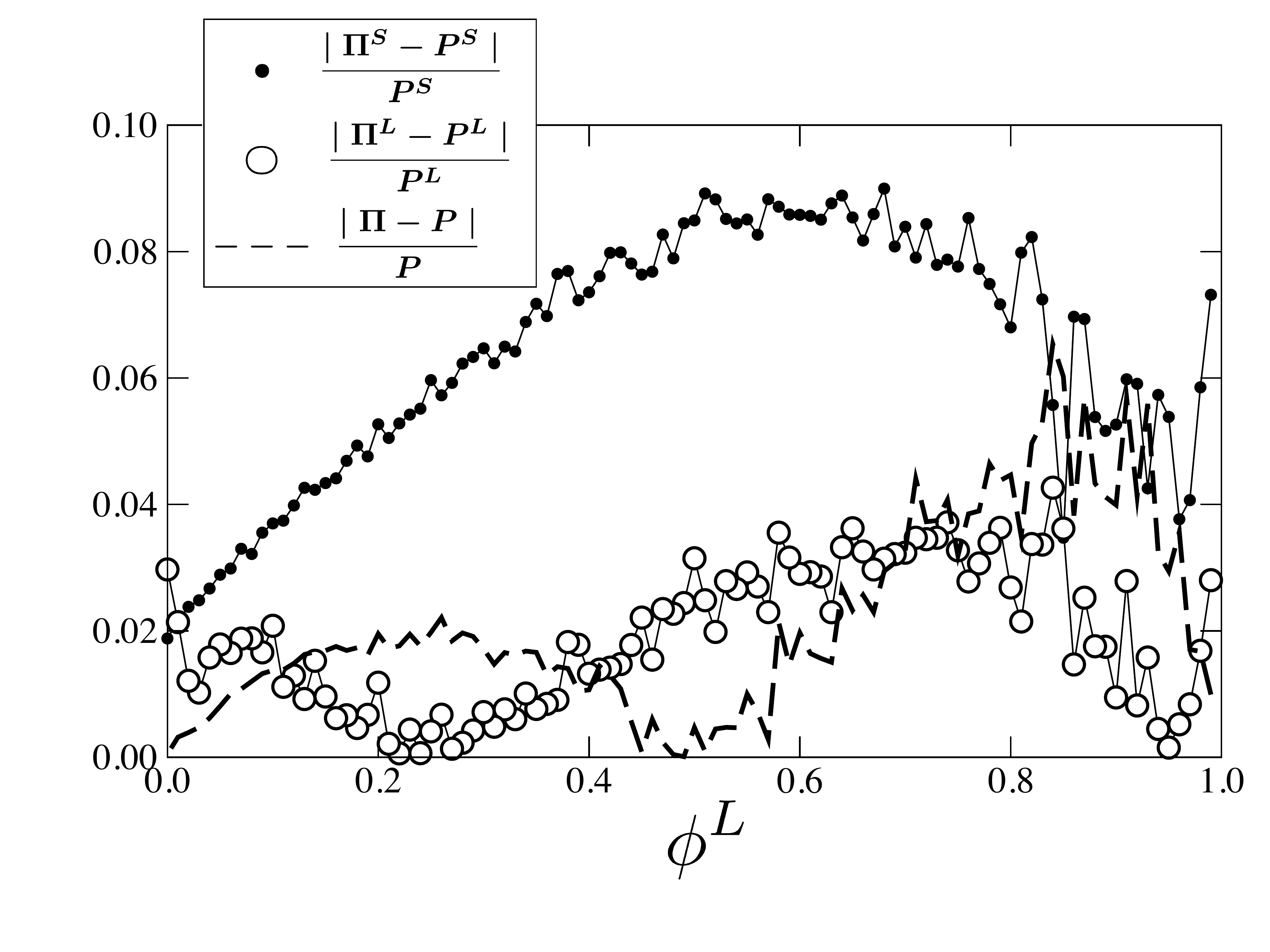}}
\caption{Mean error on the approximations $\Pi$, $\Pi^L$ and $\Pi^S$  of the pressures  $P$, $P^L$ and $P^S$  as a function of the volume fraction of large grains $\phi^L$. (For 6 flows with different volume fractions of large grains $\Phi^L=0.06$, $0.15$, $0.30$, $0.45$, $0.60$ and $0.75$,  flowing at $\theta=23^\circ$) }
\label{PIvsP}
\end{minipage}
\end{figure}
Both estimates  $\Pi^S$ and $\Pi^L$ are derived from crude simplifications and are expected to provide only a rough guess of the real value of the partial pressures as calculated from (\ref{eq:stress}). To quantify the error made when using (\ref{eq:pL}) and (\ref{eq:pS}), we compare the value they provide for $P^L$ and $P^S$ with the exact computation presented in section \ref{ContStress},  averaging  over 6 flows with different volume fractions of large grains $\Phi^L=0.06$, $0.15$, $0.30$, $0.45$, $0.60$ and $0.75$,  and flowing at an angle $\theta=23^\circ$. The results are shown in Figure \ref{PIvsP} where $\mid \Pi^S - P^S\mid/P^S$, $\mid  \Pi^L - P^L\mid/P^L$ and $\mid  \Pi - P\mid/P$ are reported as a function of $\phi_L$. We see that in spite of the simplification, $\Pi$ and $\Pi^L$  give reasonable estimates of the value of $P$ and $P^L$, with a maximum error of $3\%$ and $5\%$ respectively, and a mean error of $2.2\%$ and $1.9\%$ respectively. The error on $P^S$ is larger, but remains less than $9\%$, and has a mean value of $6.4\%$. We observe that the error is larger for large proportion of mixed contacts between large and small grains, which can be seen from Figure \ref{Partial-P}-c. As noted before, mixed contacts are the main source of uncertainty when computing partial contact stresses. \\
If the approximations are too simple to give a quantitative estimate of the stress partition coefficient $B$, they are sufficient to give a qualitative picture of how $B$ behaves with the flow micro-structure. Back to the model by Gray \& Thornton \cite{gray05}, we write  
\begin{eqnarray}
\label{bpi1} \Pi^{L} &=& \phi^L (1 + B \phi^S)\:  \Pi,\\
\label{bpi2} \Pi^{S} &=& \phi^S (1 - B \phi^L)\:   \Pi,
\end{eqnarray}
as we did for $\sigma^L_{yy}$ and   $\sigma^S_{yy}$ in section \ref{ContStress} (equations (\ref{eq:psiL}) and (\ref{eq:psiS})). Substituting $\Pi^{L}$ and $\Pi^{S}$ by their expression (\ref{eq:pL}) and (\ref{eq:pS})  gives the following for $B$:
\begin{equation}
\label{eq:B}
B \simeq \frac{ 1- r_d \frac{z_S f_S }{z_L f_L} }{ \phi^L +  r_d  \frac{z_S f_S }{z_L f_L} \times \phi^S},
\end{equation}
where $r_d = d_L/d_S = 2$. From this analysis, we see that the main ingredient that controls the contact stress partition is the ratio of the mean force resultant on small grains to the mean force resultant on large grains: the smaller this ratio, the more efficient the segregation. For $r_d=2$, the condition for segregation (namely $B>0$) is that ${z_S f^S }< \frac12{z_L f^L}$.  We note also that the size ratio $r_d$ is not explicitly favorable to stress partition, but tend to decrease the value of $B$.\\
\begin{figure}
\begin{minipage}{0.98\linewidth} 
\centerline{\includegraphics[angle=-90,width = 0.98\linewidth]{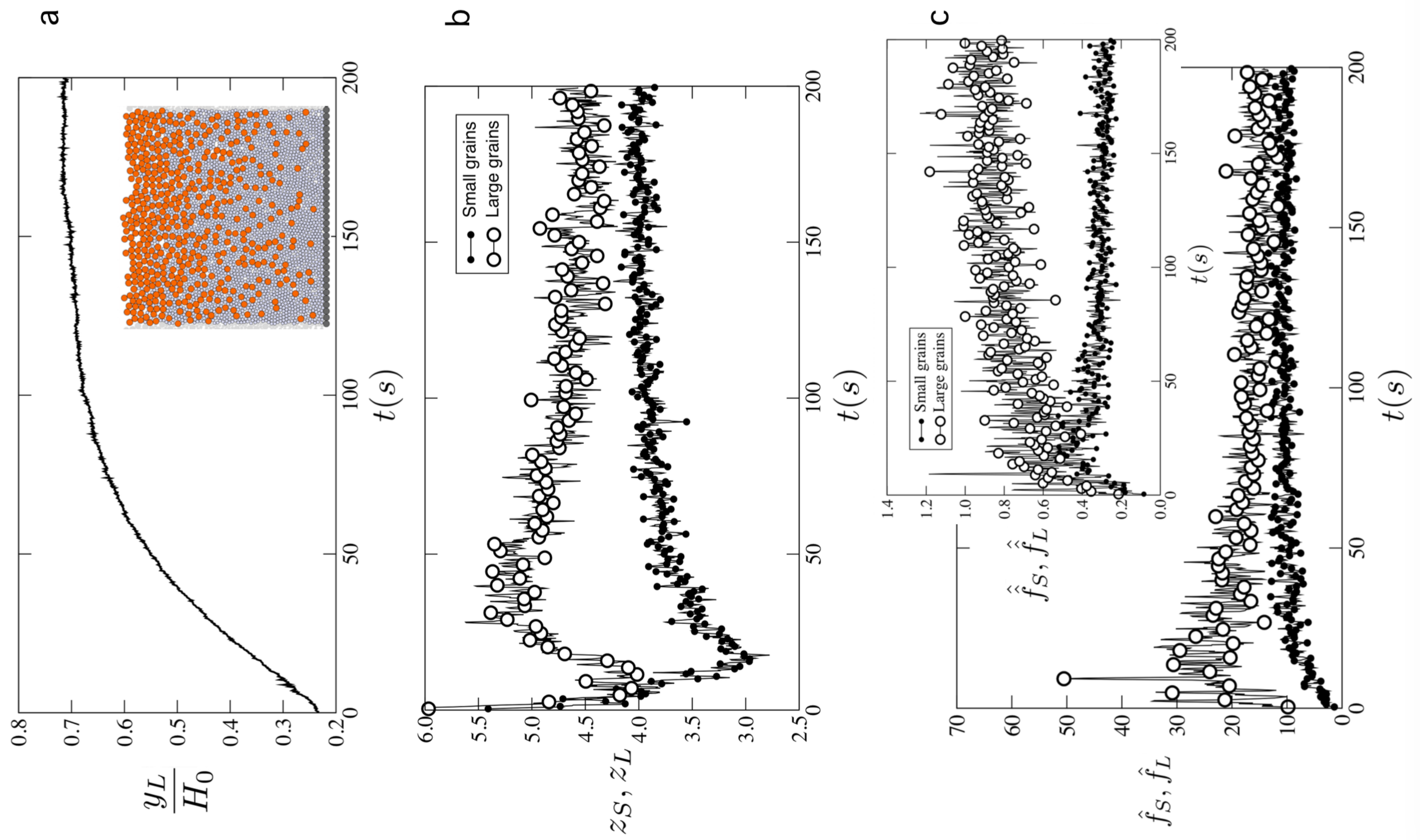}}
\caption{For a given flow with $\Phi^L = 0.45$ and $\theta=23^\circ$: {\bf a} Vertical position of the center of mass of the large grains $y_L$ normalised by the flow height $H_0$ in the course of time, {\bf b} Coordinance number for $z_S$ and $z_L$ for small ($\bullet$) and large grains ($\bigcirc$) in the course of time,  {\bf c}  Mean force transmitted by contact involving at least one small grains ($\bullet$) and one large grain  ($\bigcirc$) normalised by the weight of one small grain $m_Sg$ ($\hat{f}_S$ and $\hat{f}_L$ respectively, main graph) and  normalised by the pressure seen by their center of mass ($\hat{\hat{f}}_S$ and $\hat{\hat{f}}_L$ respectively, inset) as a function of time.}
\label{MicroStructure}
\end{minipage}
\end{figure}
This micro-structural condition for segregation can be observed from direct measurements of the mean force transmitted at contacts, and direct measurement of the mean number of contacts in which grains are involved on average for each species. 
Considering a single flow ($\Phi^L = 0.45$, $\theta=23^\circ$), we simply count  $z_S$ and $z_L$ the number of contacts in which small and large grains are involved on average as a function of time (Figure \ref{MicroStructure}-b). For the same flow, the position of the center of mass of large grains $y_L$ is  reported (Figure \ref{MicroStructure}-a). We see that the segregation process implies a significantly larger coordinance number for large grains. This is somewhat expected since  large grains can be surrounded by many small neighbours when the converse is not true. The difference between  $z_S$ and $z_L$ is maximum when the mixing is maximum. More significantly, the mean forces $f_S$ and $f_L$, transmitted by contacts involving at least one small grain and at least one large grain respectively, disclose a different behaviour for the  two phases of grains.  The normalised mean forces $\hat{f}_S = f_S/m_sg$ and  $\hat{f}_L = f_L/m_sg$ (where $m_S$ is the mass of one small grain)  are plotted in the course of time in Figure \ref{MicroStructure}-c.  At the start of the flow, the larger grains are at the bottom and are thus submitted to a larger pressure than the smaller grains closer to the surface: the fact that  $\hat{f}_S  < \hat{f}_L$ is expected. However, as segregation proceeds, and larger grains rise to the top, the inequality remains true. This effect  becomes more apparent when normalising the mean forces $f_S$ and $f_L$ by the pressure seen by the center of mass of each species ($y_S$ and $y_L$ respectively): $\hat{\hat{f}}_S = f_S/\rho g (H_0 -y_S)$ and  $\hat{\hat{f}}_L = f_L/\rho g (H_0 -y_L)$. Doing so, we filter out the influence of the grains position in the flow, and stress the fact that  forces transmitted by contacts involving large grains are significantly larger than forces transmitted at contacts involving small grains only. \\

\section{Summary and discussion}
\label{sec:conclu}

The aim of this work is to understand the size segregation mechanism in granular chute flows in terms of  pressure distribution within the two phases of grains, large and small, and to relate it to the granular micro-structure. Therefore, discrete numerical simulations of bi-disperse granular chute flows are systematically analysed. 
 Based on the theoretical models by Gray $\&$ Thornton \cite{gray05} and Hill $\&$ Tan \cite{hill14}, we compute the partial stress tensors  associated to the phases of small and large grains while separating the contribution of the contacts stresses (resulting from the forces transmitted at contacts) and the kinetic stresses (resulting from the grains velocity fluctuations). Comparing the contact pressure in the phases of small grains and large grains, we observe a slight deviation from a classical mixture, whereby small grains see less of the gravity gradient than their volume fraction implies. This result supports the possibility of a gravity driven segregation mechanism as proposed by \cite{gray05}. We show however that the contact stress partition is extremely sensitive to the definition of the partial stress tensors, and more specifically, to the way mixed contacts (namely contact involving a small grain and a large grain) are handled. As a result, the contact stress partition coefficient is sensitive to the ratio of grain sizes. This makes the interpretation of contact stress ratios awkward, particularly if one wants to compare mixtures with different grain sizes: in that case, it will be difficult to tell the bias introduced by the ponderation by the grains size in the partial stresses computation from the influence on the mixed stress partition that derives. \\
 By contrast, the computation of the partial kinetic stress tensors is more robust since the separation between the phases of large and small grains, relying on the grains themselves and not on the contacts, is straightforward. Comparing theses partial kinetic stress tensors, we find that the phase of small grains exhibits a significantly larger kinetic pressure than implied by their volume fraction. Computing a  kinetic stress partition coefficient in the spirit of \cite{gray05}, we find a value much larger than that obtained for contact stresses, and comparable to the values found in previous work by \cite{hill14} and \cite{weinhart13}. These results support the existence of a segregation mechanism induced by shear rate gradients as advocated by \cite{fan11b,hill14}, and suggest moreover that this mechanism might   be more important in amplitude than  gravity induced segregation.  In our systems, large grains are segregated towards  the cooler regions of the flow (the surface) as predicted by kinetic theory \cite{larcher13}.  In addition, we evidence an increase of the value of the kinetic stress partition coefficient with the flow shear rate, corresponding to smaller segregation time, while no such influence was observed on the contact stress partition.\\
Finally, using a simple approximation for the contact partial stress tensors, we investigate how contact stress partition relates with the flow microstructure.  We observe that grains coordinance (namely the grains average number of contacts) is greater in the phase of large grains, and that contacts involving at least one large grain transmit forces of greater amplitude than the others. \\
Our results do not allow us to decide for either gravity-induced mechanisms or shear gradient induced mechanisms, as being the most likely. However, the following points should be stressed: i) the computation of partial contact stress tensors implies a real difficulty due to the existence of mixed contacts, ii)   as a result, the analysis of partial contact stress ratios lacks robustness, iii) the analysis of the microstructure might provide a practical proxy to study gravity-driven segregation.\\
On the other hand, the robustness of kinetic stress computation, and the sensitivity of the partial kinetic stress partition to the flow state, suggest that a parametric study of these quantities varying contact properties, grains sizes and flow dynamics, would  lead to interesting new insights into segregation. \\

\noindent{\bf ACKNOWLEDGEMENT}\\
 This work was supported by the FP7 European Grant IEF n$^\circ$297843.

\end{document}